\documentclass{article}

\usepackage{amssymb,amsmath}

\newcommand{\beq}{\begin{eqnarray}}
\newcommand{\eeq}{\end{eqnarray}}
\newcommand{\beqnn}{\begin{eqnarray*}}
\newcommand{\eeqnn}{\end{eqnarray*}}
\newtheorem{theorem}{Theorem}

\newcommand{\rd}{\partial}

\newcommand{\tp}[1]{\:{}^{\mathrm{t}}#1}
\newcommand{\Ker}{\mathop{\mathrm{Ker}}}
\newcommand{\Coker}{\mathop{\mathrm{Coker}}}

\newcommand{\CC}{\mathbf{C}}
\newcommand{\PP}{\mathbf{P}}

\newcommand{\bfalpha}{\mbox{\boldmath$\alpha$}}
\newcommand{\bfbeta}{\mbox{\boldmath$\beta$}}
\newcommand{\bfgamma}{\mbox{\boldmath$\gamma$}}
\newcommand{\bfpsi}{\mbox{\boldmath$\psi$}}
\newcommand{\bfvarphi}{\mbox{\boldmath$\varphi$}}

\begin{document}

\title{Tyurin parameters of commuting pairs and \\
infinite dimensional Grassmann manifold}
\author{Kanehisa Takasaki\\
\normalsize 
Graduate School of Human and Environmental Studies,\\
\normalsize Kyoto University,\\
\normalsize Yoshida, Sakyo, Kyoto 606-8501, Japan\\
\normalsize takasaki@math.h.kyoto-u.ac.jp}
\date{}
\maketitle

\begin{abstract} 
Commuting pairs of ordinary differential operators are 
classified by a set of algebro-geometric data called 
``algebraic spectral data''.   These data consist of 
an algebraic curve (``spectral curve'') $\Gamma$ with 
a marked point $\gamma_\infty$, a holomorphic vector bundle 
$E$ on $\Gamma$ and some additional data related to 
the local structure of $\Gamma$ and $E$ in a neighborhood 
of $\gamma_\infty$.  If the rank $r$ of $E$ is greater 
than $1$,  one can use the so called ``Tyurin parameters'' 
in place of $E$ itself.  The Tyurin parameters specify 
the pole structure of a basis of joint eigenfunctions 
of the commuting pair.  These data can be translated to 
the language of an infinite dimensional Grassmann manifold.  
This leads to a dynamical system of the standard exponential 
flows on the Grassmann manifold, in which the role of 
Tyurin parameters and some other parameters is made clear. 
\end{abstract}

\section{Introduction}

My lecture at the workshop was focussed on 
the Landau-Lifshitz (LL) equation.  This equation 
is a typical soliton equation whose Lax formalism 
is based on an elliptic spectral parameter.  
My main concern is to understand this kind of 
equations in Sato's (and Segal and Wilson's) 
Grassmannian perspectives of soliton equations 
\cite{bib:SS82,bib:SW85}.  Although a huge number 
of soliton equations have been shown to fall into 
this universal picture, most of them are equations 
with a rational spectral parameter.  As regards 
the LL equation, such a Grassmannian approach 
has been achieved by Carey et al. \cite{bib:CHMS93}.  
In my lecture, I reviewed a slightly different approach 
of myself \cite{bib:Ta03a}.  Since the contents of 
the lecture overlaps with my contribution to 
the proceedings of a Faro workshop \cite{bib:Ta05}, 
I will not repeat it here.  

Another interesting class of material in this context 
can be found in Krichever's recent work  \cite{bib:Kr02} 
on the construction of Lax (and zero-curvature) equations 
defined on an arbitrary compact Riemann surface. 
These equations have the so called ``Tyurin parameters'' 
among dynamical variables.  Tyurin parameters are known 
in algebraic geometry as parameters of deformations of 
(generic) holomorphic vector bundles on an algebraic curve 
\cite{bib:Ty67}.  Krichever's Lax equations are thus related 
to deformations of holomorphic vector bundles; in contrast, 
the LL equation is associated with a rigid (though nontrivial) 
bundle.  I examined a very simple example of Krichever's 
construction, and found that the Grassmannian perspective 
is valid for this case as well \cite{bib:Ta03b}.  This result, 
too, is reviewed in the contribution to the Faro workshop 
({\it loc. cit.\/}).  

In the present article, I consider a more classical case, 
namely, commuting pairs (or commutative rings) of ordinary 
differential operators and the associated special solutions 
of the KP hierarchy.  As elucidated in the studies 
in the 1970's \cite{bib:Dr77,bib:GD75-76,bib:Kr77,
bib:Kr78,bib:KN78,bib:KN80,bib:Mu78,bib:Ve83}, 
such commuting pairs are classified by a set of 
algebro-geometric data (``algebraic spectral data'').  
These data consist of an algebraic curve (``spectral curve'') 
$\Gamma$, a holomorphic vector bundle $E$ on $\Gamma$, and 
some other additional data.  Sect. 2--5 of this article 
are devoted to a review of this subject.  The nature of 
problem drastically changes as the rank $r$ of $E$ 
exceeds $1$.  The case of $r = 1$ reduces to Jacobi's 
inversion problem, and can be solved explicitly 
by the classical theory of theta functions and 
Abelian integrals (powered by the use of Baker-Akhiezer 
functions) \cite{bib:Kr77}.  Lacking a similar theory 
for vector bundles, the case of $r > 1$ gets much harder.  
To formulate a vector version of the inversion problem, 
Krichever and Novikov \cite{bib:Kr78,bib:KN78,bib:KN80} 
employed the notion of Tyurin parameters (also referred 
to as ``matrix divisors''; see Sect. 4 and 5).  
Previato and Wilson \cite{bib:PW89} translated 
the work of Krichever and Novikov to the language 
of an infinite dimensional Grassmann manifold 
(see Sect. 6).  As I pointed out in the previous work 
\cite{bib:Ta03b}, their usage of the Grassmann manifold 
is slightly different from the usual interpretation of 
soliton equations \cite{bib:SS82,bib:SW85}. The goal 
of this article (see Sect. 7) is to show how to 
interpret this case in the usual Grassmannian 
perspective.

\section{Spectral curve}

The study of commuting pairs dates back to the beginning 
of the twentieth century.  Of particular importance 
are the pioneering works of Schur \cite{bib:Sc05} and 
Burchnall and Chaundy \cite{bib:BC23-28}; 
see Mulase's review \cite{bib:Mu-review} for 
a rather detailed historical account as well as 
a modern, scheme-theoretical interpretation of this issue.   

Although the work of Burchnall and Chaundy was done 
some twenty years afer Schur's work, let us first 
recall their work.  They pointed out that any commuting 
(i.e., $[P,Q] = 0$) pair $(Q,P)$ of ordinary differential 
operators 
\beqnn
  Q = \rd_x^m + u_2(x)\rd_x^{m-2} + \cdots + u_m(x), 
  \nonumber \\
  P = \rd_x^n + v_2(x)\rd_x^{n-2} + \cdots + v_n(x), 
\eeqnn
satisfy a polynomial relation 
\beq
  F(Q,P) = 0
\eeq
with constant coefficients --- a fact that had been known 
for a few special cases.  This implies that 
the eigenvalues of the joint eigenvalue problem 
\beqnn
  Q\psi = z\psi, \quad 
  P\psi = w\psi 
\eeqnn
satisfy the algebraic relation 
\beq
  F(z,w) = 0. 
\eeq
Roughly speaking, this equation defines the spectral curve.  

Schur's standpoint is more abstract and, in a sense, 
closer to the modern approach to this subject.  
He considered the subring 
\beq
  \mathcal{A}_Q = \{ A \in \mathcal{D} \mid [A,Q] = 0\}
\eeq
of commutants of $Q$ in the noncommutative ring $\mathcal{D}$ 
of ordinary differential operators, and observed that 
$\mathcal{A}_Q$ is a commutative ring.  From this point of view, 
the commuting pair is nothing but generators of $\mathcal{A}_Q$.  
The commutative subring $\mathcal{A}_Q \subset \mathcal{D}$ 
is a more intrinsic notion than the commuting pair $(Q,P)$.  
In the language of modern algebraic geometry, the spectral curve 
is nothing but the spectrum $\mathrm{Spec}\mathcal{A}_P$ --- 
an amusing coincidence of the usage of the word ``spectrum''. 
It is quite easy to define the rank $r$ in terms of 
$\mathcal{A}_Q$:  $r$ is the greatest common divisor of 
the orders of all operators in $\mathcal{A}_Q$.  

Krichever \cite{bib:Kr78} defined the rank in terms of 
commuting pairs.  Let us review his definition and 
its implications.  The definition consist of several steps.  

\begin{enumerate}
\item 
The first step is to consider the action of $P$ on the space 
of solutions of the ordinary differential equation 
$Q\psi = z\psi$.  This equation has an $m$-tuple 
$\varphi_k = \varphi_k(x,x_0,z)$, $k = 0,\ldots,m-1$, 
of linearly independent solutions that are normalized 
by the initial conditions $\rd_x^j\varphi_k|_{x=x_0} 
= \delta_{jk}$, $k = 0,\ldots,m-1$, at a reference point $x_0$.   
If $[P,Q] = 0$, the space of solutions of $Q\psi = z\psi$ 
is invariant under the action of $P$, so that there is 
an $m \times m$ matrix $M(x_0,z)$ such that 
\beq
  (P\varphi_0,\ldots,P\varphi_{m-1}) 
  = (\varphi_0,\ldots,\varphi_{m-1})M(x_0,z).  
\eeq
More explicitly, 
\beq
  M(x_0,z) = (\rd_x^j P \varphi_k|_{x=x_0})_{j,k=0,\ldots,m-1}. 
\eeq
The matrix elements of $M(x_0,z)$ are entire functions of $z$.  
The aforementioned polynomial $F(z,w)$ is given by 
the characteristic polynomial 
\beq
  F(z,w) = \det(wI - M(x_0,z)).  
\eeq
On the other hand, there is another $m \times m$ matrix 
$V(x,z) = (v_{jk}(x,z))$ such that 
\beq
  \rd_x^jP\psi = \sum_{k=0}^{m-1}v_{jk}(x,z)\rd_x^k\psi 
\eeq
holds for any solution of $Q\psi = z\psi$.  
The coefficients $v_{jk}(x,z)$ can be determined by 
division of differential operators, which shows 
that they are polynomials in $z$.  If one applies 
the defining relation of $v_{jk}(x,z)$ to $\varphi_j$'s 
and set $x = x_0$, one readily finds that 
\beq
  V(x_0,z) = M(x_0,z). 
\eeq
Thus $F(z,w)$ turns out to be a polynomial in 
both $z$ and $w$.  

\item 
The second step is to introduce a power series solution 
(the so called ``formal Baker-Akhiezer function'') 
\beq
  \hat{\psi}(x,x_0,\lambda) 
  = (1 + \sum_{\ell=1}^\infty \phi_\ell(x,x_0)\lambda^{-\ell}) 
    e^{(x-x_0)\lambda}
\eeq
of $Q\psi = z\psi$ 
under the initial condition $\hat{\psi}(x_0,x_0,\lambda) = 1$. 
The parameter $\lambda$ is related to $z$ as 
\beq
  z = \lambda^m. 
\eeq
The action of $P$ on $\hat{\psi}(x,x_0,\lambda)$ defines 
a Laurent series $p(\lambda) = \lambda^n + \cdots$ with 
constant coefficients as 
\beq
  P\hat{\psi}(x,x_0,\lambda) 
  = p(\lambda)\hat{\psi}(x,x_0,\lambda). 
\eeq
Replacing $\lambda \to e^{2\pi ik/m}\lambda$ yields 
\beqnn
  P\hat{\psi}(x,x_0,e^{2\pi ik/m}\lambda) 
  = p(e^{2\pi ik/m}\lambda)
    \hat{\psi}(x,x_0,e^{2\pi ik/m}\lambda). 
\eeqnn
Thus we have an $m$-tuple of solutions 
$\hat{\psi}(x,x_0,e^{2\pi ik/m}\lambda)$, 
$k = 0,\ldots,m-1$, to the equation $Q\psi = z\psi$ 
on which the action of $P$ diagonalizes.  
Consequently, the characteristic polynomial $F(z,w)$ 
of $M(x_0,z)$ factorizes as 
\beq
  F(z,w) = \prod_{j=0}^{m-1} (w - p(e^{2\pi ik/m}\lambda)). 
\eeq
Note that this is an equality that holds in a neighborhood 
of $z = \infty$.  

\item 
Let $\tilde{m}$ be the smallest positive integer 
for which $p(e^{2\pi ik/m}\lambda)$, $k = 1,2,\ldots$, 
returns to $p(\lambda)$, i.e., $p(e^{2\pi ik/m}\lambda) 
\not= p(\lambda)$ for $k=1,\ldots,\tilde{m}-1$ 
and $p(e^{2\pi i\tilde{m}/m}\lambda) = p(\lambda)$.  
The rank $r$ is now defined by 
\beq
  r = m/\tilde{m}. 
\eeq
The Laurent series $p(\lambda)$ can be written as 
\beq
  p(\lambda) = \tilde{p}(\lambda^r) 
\eeq
with another Laurent series $\tilde{p}(\kappa) 
= \kappa^{\tilde{n}} + \cdots$ ($\tilde{n} = n/r$) of 
\beq
  \kappa = \lambda^r. 
\eeq
Since $p(e^{2\pi ik/m}\lambda)$ is $\tilde{m}$-periodic 
with respect to $k$, the foregoing (local) expression 
of $F(z,w)$ factorizes as 
\beq
  F(z,w) = f(z,w)^r, 
\eeq
where 
\beq
  f(z,w) 
  = \prod_{j=0}^{\tilde{m}-1}
    (w - \tilde{p}(e^{2\pi ij/\tilde{m}}\lambda^r)). 
\eeq
By construction, $f(z,w)$ is single-valued in 
a neighborhood of $z = \infty$, hence becomes 
a polynomial in $z$ as well.  
\end{enumerate}

The equation $F(z,w) = 0$ thus turns out 
to be reducible.  We define the (affine) 
pectral curve by the reduced equation 
\beq
  f(z,w) = 0. 
\eeq
Since the branches of the solutions of $f(z,w) = 0$ 
in a neighborhood of $z = \infty$ are parameterized as 
\beqnn
  (z,w) = (\kappa^{\tilde{m}},
           \tilde{p}(e^{2\pi ik/\tilde{m}}\kappa)), 
  \quad k = 0,\ldots,\tilde{m}-1, 
\eeqnn
this curve can be compactified by adding a point 
$\gamma_\infty$ at infinity, $\kappa^{-1}$ being 
a local coordinate in a neighborhood of 
$\gamma_\infty$.  Let $\Gamma$ denote 
the compactified spectral curve, and $\Gamma_0$ 
the affine part $\Gamma\setminus\{\gamma_\infty\}$.

\section{Holomorphic vector bundle}

Let $\bfvarphi = \bfvarphi(x,x_0,z)$ denote the row vector 
\beqnn
  \bfvarphi 
  = (\varphi_0,\ldots,\varphi_{m-1}) 
\eeqnn
of the aforementioned fundamental solutions of 
$Q\psi = z\psi$.  If $\mathbf{c}$ is an eigenvector 
of $M(x_0,z)$ with eigenvalue $w$, $\psi 
= \bfvarphi\mathbf{c}$ gives a joint eigenfunction 
with spectrum $(z,w)$. The fact that $F(z,w)$ 
factorizes to $f(z,w)^r$ means that each eigenvalue 
of $M(x_0,z)$ is $r$-fold degenerate and that 
the eigenspace is $r$-fold degenerate.  If we choose 
a basis $\mathbf{c}_0,\ldots,\mathbf{c}_{r-1}$ 
of the eigenspace of $M(x_0,z)$, the associated 
joint eigenfunctions $\psi_k = \bfvarphi\mathbf{c}_k$,  
$k = 0,\ldots,r-1$, form a basis of the space of 
joint eigenfunctions 
\beq
  E_{(z,w)} = \{\psi \mid Q\psi = z\psi,\; P\psi = w\psi\}. 
\eeq
Putting this vector space at each point $(z,w)$, 
we obtain a holomorphic vector bundle $E_0$ of rank $r$ 
on the affine part $\Gamma_0$ of the spectral curve. 
\footnote{Actually, the situation gets complicated 
when the equation $f(z,w) = 0$ for $w$ has a multiple 
root, namely, when $(z,w)$ is a branch point of 
the covering of $\pi: \Gamma_0 \to \CC\PP^1$, 
$\pi(z,w) = z$.  A careful analysis shows that 
the joint eigenfunctions persists to be holomorphic 
functions in a neighborhood of those branch points, 
so that $E_0$ is indeed a holomorphic vector bundle 
over the whole $\Gamma_0$.}  

This bundle $E_0$ can be extended to a holomorphic 
vector bundle $E$ on the compactified spectral curve 
$\Gamma$.  This is the place where we find 
a final piece of geometric data, namely, the choice 
of local trivialization of $E$ in a neighborhood of 
$\gamma_\infty$ \cite{bib:PW89}.  In the case of $r = 1$, 
this part of the geometric data is less important 
(even negligible).  In the case of $r > 1$, in contrast, 
the choice of local trivialization of $E$ plays 
a substantial role, as one can see in the work of 
Li and Mulase \cite{bib:Mu90,bib:LM97,bib:Mu-review}.  

Actually, Krichever and Novikov 
\cite{bib:Kr78,bib:KN78,bib:KN80} introduce 
an alternative set of data here.  This data consist 
of $r-1$ functions $w_2(x),\ldots,w_r(x)$ (hence absent 
if $r = 1$), and related to the asymptotic behavior of 
the joint eigenfunctions $\psi_k$ as $z \to \infty$.  
Let us normalize these joint eigenfunctions by 
the initial conditions 
\beq
  \rd_x^j\psi_k|_{x=x_0} = \delta_{jk}, 
  \quad k = 0,\ldots,r-1. 
\eeq
As $z \to \infty$, these joint eigenfunctions 
behave as 
\beq
  (\psi_0,\ldots,\psi_{r-1}) 
  = (\xi_0,\ldots,\xi_{r-1})\Psi_0, 
  \label{eq:bfpsi-Psi0}
\eeq
where $\xi_0,\ldots,\xi_{r-1}$ are Laurent series 
of $\kappa$ of the form 
\beq
  \xi_k = \delta_{k,0} 
    + \sum_{\ell=1}^\infty \xi_{k\ell}\kappa^{-\ell}, 
\eeq
and $\Psi_0 = \Psi_0(x,x_0,\kappa)$ is a solution 
of the matrix differential equation 
\beq
  \rd_x\Psi_0 = A_0 \Psi_0 
\eeq
with the coefficient matrix 
\beqnn
  A_0 
  = \left(\begin{array}{ccccc}
      0     & 1    & 0    &\cdots& 0 \\
      0     & 0    & 1    &\ddots&\vdots \\
      \vdots&\vdots&\ddots&\ddots& 0 \\
      0     & 0    &\cdots&0     & 1 \\
      \kappa-w_r(x)&-w_{r-1}(x)&\cdots&-w_2(x)& 0 
    \end{array}\right) 
\eeqnn
normalized by the initial condition 
\beq
  \Psi_0|_{x=x_0} = I. 
\eeq
Of course this matrix system is equivalent to 
the scalar equation 
\beq
  (\rd_x^r + w_2(x)\rd_x^{r-2} + \cdots + w_r(x))\psi = 0. 
\eeq
The entries of the first row of $\Psi_0$ are 
a set of fundamental solutions of this equation.  
The $r-1$ functions $w_2(x),\ldots,w_r(x)$ are 
the final data that Krichever and Novikov use 
in their approach to commuting pairs of differential 
operators.  

We can now extend the bundle $E_0$ over $\Gamma_0$ 
to a bundle $E$ over $\Gamma$ using $\Psi_0$ 
as the transition function.  $\xi_k$'s are 
interpreted as a basis of holomorphic sections 
of $E$ in a neighborhood of $\gamma_\infty$, 
hence determines local trivialization of $E$ 
therein.

\section{Tyurin parameters}

The normalized joint eigenvectors $\psi_k$ are 
expressed as $\psi_k = \bfvarphi\mathbf{c}_k$, 
where $\mathbf{c}_k = (c_{jk})_{j=0,\ldots,m-1}$ 
are eigenvectors of $M(x_0,z)$ normalized as 
\beq
  c_{jk} = \delta_{jk}, 
  \quad j,k = 0,\ldots,r-1. 
\eeq
These normalization conditions uniquely 
determine the eigenvectors $\mathbf{c}_k$, 
which thereby become meromorphic functions 
$\mathbf{c}_k(x_0,\gamma)$ of $\gamma = (z,w)$ 
on $\Gamma_0$.  Let $\gamma_s$, $s = 1,\ldots,N$, 
denote the poles of $\mathbf{c}_k$'s.  
Since the components of $\bfvarphi$ are 
entire functions of $z$, the joint eigenfunctions 
$\psi_k$, too, are meromorphic functions 
$\psi_k(x,x_0,\gamma)$ of $\gamma$ on $\Gamma_0$ 
with poles at $\gamma_s$, $s = 1,\ldots,N$.  

According to Krichever \cite{bib:Kr78}, 
the multiplicity $m_s$ of these poles $\gamma_s$ 
satisfy the equality 
\beq
  \sum_{s=1}^N m_s = rg. 
\eeq
In a generic situation, these poles are all simple 
(i.e., $m_s = 1$) so that $\psi_k$'s have $rg$ 
simple poles $\gamma_1,\ldots,\gamma_{rg}$ 
in addition to an essential singularity at 
$\gamma_\infty$.  In the following, we assume 
this generic situation.  

In the case of $r = 1$ where $E$ is a line bundle, 
the joint eigenfunction is nothing but the usual 
Baker-Akhiezer function \cite{bib:Kr77}.  
Such a scalar Baker-Akhezer function is uniquely 
determined by the asymptotic behavior in 
a neighborhood of $\gamma_\infty$ and the position 
of the $g$ poles $\gamma_1,\ldots,\gamma_g$ or, 
rather, by the divisor $\gamma_1 + \cdots + \gamma_g$.  
This divisor, in turn, determines the line bundle $E$.  

If $r > 1$, the divisor $\gamma_1 + \cdots + \gamma_{rg}$ 
is not enough to specify the vector bundle $E$.  
To overcome this difficulty, Krichever and Novikov 
employ the notion of ``matrix divisors'' (i.e., 
Tyurin parameters) in the sense of Tyurin \cite{bib:Ty67}.  
Let $\bfpsi$ denote the row vector 
\beqnn
  \bfpsi(x) = (\psi_0,\ldots,\psi_{r-1})
\eeqnn
of the joint eigenfunctions.  This vector-valued 
meromorphic function on $\Gamma_0$ 
(``vector Baker-Akhiezer function'' 
in the terminology of Krichever and Novikov 
\cite{bib:KN78,bib:KN80}) have simple poles 
at $\gamma_1,\ldots,\gamma_{rg}$.  
The Tyurin parameters $(\gamma_s,\bfalpha_s)$, 
$s = 1,\ldots,rg$, consist of the position of 
these poles $\gamma_s$ and the complex ray 
$\bfalpha_s \in \PP^{r-1}$ 
determined by the residue of $\bfpsi$ at $\gamma_s$.  
\footnote{Precisely speaking, this is slightly 
different from the usage of Tyurin parameters 
in algebraic geometry. \cite{bib:Ty67}.  Namely, 
$(\gamma_s,\bfalpha_s)$'s are Tyurin parameters 
of the dual bundle $E^*$ of $E$ \cite{bib:PW89}.} 
One can normalize the directional vectors 
$\bfalpha_s$ as 
\beqnn
  \bfalpha_s = (\alpha_{s,0},\ldots,\alpha_{s,r-2},1). 
\eeqnn
$\bfalpha_s$ arises in the local expression 
of $\bfpsi$ in a neighborhood of $\gamma_s$ as 
\beq
  \bfpsi
  = \frac{\beta_s \bfalpha_s}{z - z(\gamma_s)} 
    + O(1), 
  \label{eq:bfpsi-at-gamma}
\eeq
where $z(\gamma_s)$ denotes the $z$-coordinate of 
$\gamma_s$, and $\beta_s$ is a scalar factor.  

We have thus obtained the algebraic spectral data 
\beqnn
  \Sigma = 
    (\Gamma,\gamma_\infty,\kappa, 
    (\gamma_s,\bfalpha_s)_{s=1}^g, 
    (w_j(x))_{j=2}^r)
\eeqnn
of the commuting pair $(Q,P)$.  These are 
an analogue of the ``scattering data'' in 
the inverse scattering problem.  
The inverse problem reduces to a kind of 
Riemann-Hilbert problem, namely, 
to find from $\Sigma$ a vector-valued analytic 
function $\bfpsi$ on $\Gamma$ that has simple poles 
at $\gamma_1,\ldots,\gamma_{rg}$ and 
essential singularity at $\gamma_\infty$, 
and behaves as (\ref{eq:bfpsi-Psi0}) and 
(\ref{eq:bfpsi-at-gamma}) in a neighborhood of 
these singular points.  Krichever \cite{bib:Kr78} 
solved  this problem by the standard method of 
integral equation with a Cauchy kernel on $\Gamma$. 
Previato and Wilson \cite{bib:PW89} reformulated 
Krichever's method in the language of 
an infinite dimensional Grassmann manifold.

\section{Another set of Tyurin parameters}

Alongside the foregoing Tyurin parameters, 
there is  another set of Tyurin parameters that 
amounts to the divisor of {\it zeroes} of 
the usual Baker-Akhiezer function.  To distinguish 
between these two sets of parameters, let 
$(\gamma_s(x_0),\bfalpha_s(x_0))$, $s = 1,\ldots,rg$, 
denote the previous Tyurin parameters (because 
they depend on the reference point $x_0$), and 
$(\gamma_s(x),\bfalpha_s(x))$, $s - 1,\ldots,rg$, 
the second set of Tyurin parameters.  As the notation 
implies, they do depend on $x$ and coalesce to 
$(\gamma_s(x_0),\bfalpha_s(x_0))$ as $x \to x_0$.  

To introduce the second set of Tyurin parameters, 
let us consider the Wronskian matrix 
\beq
  \Psi 
  = \left(\begin{array}{c}
    \bfpsi \\
    \rd_x\bfpsi \\
    \vdots \\
    \rd_x^{r-1}\bfpsi 
    \end{array}\right) 
  = \left(\begin{array}{ccc}
    \psi_0 & \cdots & \psi_{r-1} \\
    \rd_x\psi_0 & \cdots & \rd_x\psi_{r-1} \\
                & \cdots &  \\
    \rd_x^{r-1}\psi_0 & \cdots & \rd_x^{r-1}\psi_{r-1} 
    \end{array}\right) 
\eeq
of $\psi_k$'s.  By (\ref{eq:bfpsi-Psi0}), 
this matrix-valued function can be expressed as 
\beq
  \Psi = \Xi \Psi_0(x,x_0,\kappa), \quad 
  \Xi = \sum_{\ell=0}^\infty \Xi_\ell\kappa^{-\ell}, 
\eeq
in a neighborhood of $\gamma_\infty$.  The leading 
coefficient $\Xi_0$ is a lower triangular matrix 
whose diagonal elements are all equal to $1$.  
Moreover, $\Psi_0$ is unimodular (i.e., 
$\det\Psi_0 = 1$) because the coefficient matrix 
of the differential equation for of $\Psi_0$ 
is trace-free.  Consequently, $\det\Psi$ is 
no longer singular at $\gamma_\infty$ but 
behaves as $\det\Psi = 1 + O(\kappa^{-1})$.  
Thus $\det\Psi$ turns out to be a meromorphic 
function on the whole spectral curve $\Gamma$.  

$\det\Psi$ has poles at $\gamma_1(x_0),\ldots,
\gamma_{rg}(x_0)$ and is holomorphic elsewhere.  
Since $\bfpsi$ beheaves as 
\beqnn
  \bfpsi 
  = \frac{\beta_s(x)\bfalpha_s(x_0)}{z - z(\gamma_s(x_0)} 
    + O(1) 
\eeqnn
in a neighborhood of $\gamma_s(x_0)$, 
the residue of the Wronskian matrix $\Psi$ 
turns out to be a rank-one matrix: 
\beq
  \Psi 
  = \frac{\tp{\bfbeta_s(x)}\bfalpha_s(x_0)}
    {z - z(\gamma_s(x_0))} 
    + O(1), 
\eeq
where $\bfbeta_s(x)$ is a vector-valued function.  
This implies that $\det\Psi$ has at most a simple 
pole at $\gamma_s(x_0)$.  Since $\det\Psi \to 1$ 
as $x \to x_0$, the zeroes of $\det\Psi$ coalesce 
to the poles $\gamma_1(x_0),\ldots,\gamma_{rg}(x_0)$.  
Thus $\det\Psi$ turns out to have, generically, 
$rg$ simple zeroes $\gamma_s(x)$ that tends 
to $\gamma_s(x_0)$ as $x \to x_0$.  

The second set of Tyurin parameters consist of 
the pairs $(\gamma_s(x),\bfalpha_s(x))$, 
$s = 1,\ldots,rg$.  Since $\det\Psi$ has a zero 
at $\gamma_s(x)$, $\Psi^{-1}$ has a pole there.  
If the rank of the residue matrix is greater 
than $1$, $\det\Psi^{-1}$ has a multiple pole 
there --- a contradiction.  Thus the residue 
of $\Psi^{-1}$, too, turns out to be 
a rank-one matrix: 
\beq
  \Psi^{-1} 
  = \frac{\tp{\tilde{\bfbeta}_s(x)}\bfalpha_s(x)}
    {z - z(\gamma_s(x))} 
    + O(1). 
\eeq

When Krichever and Novikov \cite{bib:KN78,bib:KN80} 
introduced these $x$-dependent Tyurin parameters, 
they derived these parameters from 
the coefficient matrix $A = A(x,x_0,\gamma)$
of the matrix differential equation 
\beq
  \rd_x\Psi = A\Psi
\eeq
satisfied by $\Psi$.  Because of 
the Wronskian structure, $A$ becomes 
a matrix of the form 
\beqnn
  A 
  = \left(\begin{array}{ccccc}
      0     & 1    & 0    &\cdots& 0 \\
      0     & 0    & 1    &\ddots&\vdots \\
      \vdots&\vdots&\ddots&\ddots& 0 \\
      0     & 0    &\cdots&0     & 1 \\
      -a_r  &-a_{r-1}&\cdots&-a_2& -a_1  
    \end{array}\right), 
\eeqnn
where $a_j$ are meromorphic functions 
$a_j(x,x_0,\gamma)$ of $\gamma \in \Gamma$ with 
poles at $\gamma_1(x),\ldots,\gamma_{rg}(x)$ 
and $\gamma_\infty$.  Note that 
the matrix differential equation 
is equivalent to the scalar equations 
\beq
  (\rd_x^r + a_1\rd_x^{r-1} + \cdots + a_r)\psi_k = 0 
\eeq
for the joint eigenfunctions.  As Previato 
and Wilson pointed out \cite{bib:PW92}, 
the scalar differential operator $G 
= \rd_x^r + a_1\rd_x^{r-1} + \cdots + a_r$ 
is nothing but the (noncommutative) 
greatest common divisor of $Q - z$ and $P - w$, 
which can be calculated by Euclidean division 
of ordinary differential operators.  

These $x$-dependent Tyurin parameters 
$(\gamma_s(x),\bfalpha_s(x))$, $s = 1,\ldots,rg$, 
play the role of dynamical variables.  
If $Q$ and $P$ obey the time evolutions 
\beq
  \frac{\rd Q}{\rd t_k} = [B_k,Q], \quad 
  \frac{\rd P}{\rd t_k} = [B_k,P], \quad 
  B_k = (Q^{k/m})_{+}, 
\eeq
of the KP hierarchy, the Tyurin parameters 
also depend on the time variables 
$t = (t_1,t_2,\ldots)$; 
let $(\gamma_s(x,t),\bfalpha_s(x,t))$, 
$s = 1,\ldots,rg$, denote those time-dependent 
Tyurin parameters.  Following Krichever and Novikov 
\cite{bib:KN78,bib:KN80}, one can reformulate 
the KP hierarchy to zero-curvature equations 
\beq
  [\rd_{t_k} - A_k(x,t,\gamma),\, 
   \rd_x - A(x,t,\gamma)] = 0, \nonumber \\{}
  [\rd_{t_j} - A_j(x,t,\gamma),\, 
   \rd_{t_k} - A_k(x,t,\gamma)] = 0 
\eeq
for $r \times r$ matrices $A(x,t,\gamma)$ and 
$A_k(x,t,\gamma)$ of meromorphic functions 
on $\Gamma$.  These matrices have poles 
at $\gamma_1(x,t),\ldots,\gamma_{rg}(x,t)$ 
and $\gamma_\infty$, and are holomorphic 
elsewhere.  The poles of $A_k(x,t,\gamma)$ 
at $\gamma_s(x,t)$'s are simple, and exhibit 
the same rank-one structure as $\Psi^{-1}$.  
The zero-curvature equations are accompanied 
by a set of linear differential equations 
\beq
  \rd_x\Psi = A(x,t,\gamma)\Psi, \quad 
  \rd_{t_k}\Psi 
    = A_k(x,t,\gamma)\Psi - \Psi M_k(t,\gamma), 
\eeq
where $M_k(t,\gamma)$ are matrices that do not 
depend on $x$ (but can depend on $t$).  
One can eliminate these extra terms by 
the right gauge transformation $\Psi \to 
\Psi C(t,\gamma)$ with a suitable matrix 
$C(t,\gamma)$ (details are omitted here).

\section{Infinite dimensional Grassmann manifold}

Although one can formulate the subsequent results 
in the functional analytic framework of Segal and 
Wilson \cite{bib:SW85} as well, let us use 
the algebraic or complex analytic language of 
Sato and Sato \cite{bib:SS82}.  

An algebraic model the relevant Grassmann manifold 
is based on the vector space 
\beqnn
  V^{\mathrm{alg}} 
  = \CC((\kappa))^{\oplus r} 
  \simeq \CC((\lambda)) 
\eeqnn
of $r$-component row vectors $\mathbf{f} 
= (f_0(\kappa),\ldots,f_{r-1}(\kappa))$ of 
formal Laurent series of $\kappa$.  The isomorphism 
to $\CC((\lambda))$ is given by the mapping 
\beqnn
  (f_0(\kappa),\ldots,f_{r-1}(\kappa)) 
  \longmapsto \sum_{j=0}^{r-1}f_j(\lambda^r)\lambda^j. 
\eeqnn
Let $V^{\mathrm{alg}}_{-}$ denote the vector subspace 
\beqnn
  V^{\mathrm{alg}}_{-} 
  = \kappa^{-1}\CC[[\kappa^{-1}]]^{\oplus r} 
  \simeq \lambda^{-1}\CC[[\lambda^{-1}]]. 
\eeqnn
The Grassmann manifold $\mathrm{Gr}(V^{\mathrm{alg}})$ 
consists of closed (with respect to a topology 
of $V^{\mathrm{alg}}$) vector subspaces $W 
\subset V^{\mathrm{alg}}$ such that 
\beqnn
    \dim\Ker(W \to V^{\mathrm{alg}}/V^{\mathrm{alg}}_{-}) 
  = \dim\Coker(W \to V^{\mathrm{alg}}/V^{\mathrm{alg}}_{-}) 
  < \infty, 
\eeqnn
where $W \to V^{\mathrm{alg}}/V^{\mathrm{alg}}_{-}$ 
denote the composition of the inclusion $W \hookrightarrow 
V^{\mathrm{alg}}$ and the projection $V^{\mathrm{alg}} 
\to V^{\mathrm{alg}}/V^{\mathrm{alg}}_{-}$.  

Actually, we need an {\it analytic} version of this model.  
The analytic model is based on the vector space $V$ 
of vector-valued Laurent series 
\beq
  \mathbf{f} 
  = (f_0,\ldots,f_{r-1}) 
  = \sum_{\ell=-\infty}^\infty \mathbf{f}_\ell\kappa^\ell 
\eeq
that converge in a neighborhood (which may depend on 
$\mathbf{f}$) of $\kappa = \infty$ except at the center.  
The Grassmann manifold $\mathrm{Gr}(V)$ is defined 
in the same manner as the algebraic model except 
that $V^{\mathrm{alg}}$ is replaced by $V_{-} 
= V^{\mathrm{alg}} \cap V$.  Namely, a point of 
$\mathrm{Gr}(V)$ is represented by a closed 
vector subspace $W \subset V$ such that 
\beq
    \dim\Ker(W \to V/V_{-}) 
  = \dim\Coker(W \to V/V_{-}) 
  < \infty. 
\eeq
Let $\mathrm{Gr}^\circ(V)$ denote the so called 
``big cell'', namely, the subset that consist of 
$W$'s for which 
\beq
  W \simeq V/V_{-}. 
\eeq
Such a subspace has a basis $\{\mathbf{w}_{n,j} \mid 
n = 0,1,2,\ldots,\; j = 0,\ldots,r-1\}$ of the form 
\beq
  \mathbf{w}_{n,j} 
  = \mathbf{e}_j\kappa^n + O(\kappa^{-1}), 
\eeq
where $\mathbf{e}_j$, $j = 0,\ldots,r-1$, 
denote the standard basis of $\CC^r$: 
$\mathbf{e}_j = (\delta_{jk})_{k=0,\ldots,r-1}$.

\section{Dressed vacua in Grassmann manifold}

Following Previato and Wilson \cite{bib:PW89}, 
we now consider a special point 
$W_0(\bfgamma,\bfalpha)$ of $\mathrm{Gr}^\circ(V)$ 
for a given set of (constant) Tyurin parameters 
$(\bfgamma,\bfalpha) 
= (\gamma_s,\bfalpha_s)_{s=1}^{rg}$.  
$W_0(\bfgamma,\bfalpha)$ consists of $r$-tuples 
$\mathbf{f} = (f_0,\ldots,f_{r-1})$ as follows: 
\begin{enumerate}
\item 
$f_s = f_s(\gamma)$, $s = 0,\ldots,r-1$, 
are meromorphic functions on the affine part 
$\Gamma_0$ of the spectral curve with possible 
poles at $\gamma_1,\ldots,\gamma_{rg}$.  
$\mathbf{f}$ is identified with an elements of 
$V$ by Laurent expansion at $\gamma_\infty$.  
It is here that the local geometric data 
$(\gamma_\infty,\kappa)$ play a role.  
\item 
All poles at $\gamma_s$'s are simple.  
As $\gamma \to \gamma_s$, $\mathbf{f}$ behaves as 
\beq
  \mathbf{f} 
  = \frac{\beta_s\bfalpha_s}{z - z(\gamma_s)} 
    + O(1), 
\eeq
where $\beta_s$ is a constant scalar.  
\end{enumerate}
One can show, by the Riemann-Roch theorem, 
that $W_0(\bfgamma,\bfalpha)$ has a basis 
$\{\mathbf{w}_{n,j} \mid n = 0,1,2,\ldots,\; 
j = 0,\ldots,r-1\}$ of the form mentioned above. 
 
Previato and Wilson use this special point 
of $\mathrm{Gr}^\circ(V)$ to reformulate 
the inverse construction of Krichever and 
Novikov.   Of course, $\bfgamma$ and $\bfalpha$ 
are identified with the Tyurin parameters 
in the algebraic spectral data.  
The functional data $w_2,\ldots,w_r$ are 
encoded into the matrix-valued function $\Psi_0$. 
To combine these data, Previato and Wilson 
consider the subspace 
\beq
  W = W_0(\bfgamma,\bfalpha)\Psi_0^{-1} 
\eeq
of $V$.  This is a Grassmannian version of 
``dressing'' that is commonly used in 
the theory of integrable systems; 
$W_0(\bfgamma,\bfalpha)$ plays the role 
of ``vacuum''.  A clue here is the fact 
that $\mathrm{Gr}^\circ(V)$ is an open 
subset of $\mathrm{Gr}(V)$.  Because of 
this fact, $W$ remains in $\mathrm{Gr}^\circ(V)$ 
as far as $\Psi_0$ is sufficiently close to 
the unit matrix (this is indeed the case 
if $x$ is close to $x_0$, because 
$\Psi_0$ satisfies the initial condition 
$\Psi_0|_{x=x_0} = I$).  If $W$ indeed 
belongs to $\mathrm{Gr}^\circ(V)$, then 
one obtains an element $\bfpsi$ of $W$ 
as the inverse image of $\mathbf{e}_0$ 
by the isomorphism $W \to V/V_{-}$.  
This $\bfpsi$ gives a solution to 
the Riemann-Hilbert problem in Krichever 
and Novikov's inverse construction.  

Our usage of $W_0(\bfgamma,\bfalpha)$ 
is similar, but conceptually different.  
Namely, in place of $\Psi_0$, we take 
a matrix $\Phi$ of Laurent series of the form 
\beq
  \Phi = \sum_{\ell=0}^\infty \Phi_\ell\kappa^{-\ell}, 
\eeq
where $\Phi_0$ is a lower triangular matrix 
whose diagonal elements are equal to $1$, 
and ``dress'' $W_0(\bfgamma,\bfalpha)$ 
with $\Phi$ as 
\beq
  W = W_0(\bfgamma,\bfalpha)\Phi. 
\eeq
We use this matrix $\Phi$ to encode the data of 
local trivialization of $E$ in a neighborhood 
of $\gamma_\infty$.  

The information of local trivialization of 
$E$ is carried by the formal Baker-Akhiezer 
function $\hat{\psi} = \hat{\psi}(x,x_0,\lambda)$ 
as well as $w_2(x),\ldots,w_r(x)$.  To extract it, 
we split $\hat{\psi}$ into an $r$-tuple 
$\hat{\psi}_j = \hat{\psi}_j(x,x_0,\kappa)$, 
$j = 0,\ldots,r-1$, of power series of $\kappa$ as 
\beq
  \hat{\psi} 
  = \sum_{j=0}^{r-1}\hat{\psi}_j(x,x_0,\kappa)\lambda^j
\eeq
and construct the Wronskian matrix 
\beq
  \hat{\Psi} 
  = \left(\begin{array}{ccc}
    \hat{\psi}_0 &\cdots& \hat{\psi}_{r-1}\\
    \rd_x\hat{\psi}_0 &\cdots& \rd_x\hat{\psi}_{r-1}\\
          &\cdots& \\
    \rd_x^{r-1}\hat{\psi}_0 
          &\cdots& \rd_x^{r-1}\hat{\psi}_{r-1}
    \end{array}\right). 
\eeq
By construction, $\hat{\Psi}$ is a mattrix-valued 
Laurent series of the form 
\beq
  \hat{\Psi} = \Phi \exp((x-x_0)\Lambda(\kappa)), 
\eeq
where 
\beqnn
  \Lambda(\kappa) 
  = \left(\begin{array}{cccc}
    0 & 1      &        &   \\
      & \ddots & \ddots &   \\
      &        & \ddots & 1 \\
    \kappa &   &        & 0 
    \end{array}\right). 
\eeqnn
The first factor $\Phi = \Phi(x,x_0,\kappa)$ 
on the right hand side is a matrix-valued 
Laurent series of the form mentioned above.  
We use this $\Phi$ to ``dress'' the ``vacuum''  
$W_0(\bfgamma,\bfalpha)$.

\section{Dynamical system on Grassmann manifold}

We are now in a position to formulate our Grassmannian 
perspective of rank-$r$ commuting pairs (and 
the associated solutions of the KP hierarchy).  

To this end, we start from the fact that  
$\Psi = \Psi(x,x_0,\gamma)$ and 
$\hat{\Psi} = \hat{\Psi}(x,x_0,\kappa)$ 
satisfy the same differential equation 
\beqnn
  \rd_x\Psi = A\Psi, \quad 
  \rd_x\hat{\Psi} = A\hat{\Psi}. 
\eeqnn
(This is a consequence of the fact that 
both $\psi_j$ and $\hat{\psi}$ are joint 
eigenfunctions of $Q$ and $P$.)  
This implies that the ``matrix ratio'' 
$\Psi^{-1}\hat{\Psi}$ is independent of $x$, 
which is equal to the initial value 
$\hat{\Psi}(x_0,x_0,\kappa)$.  (Note that 
the initial value of $\hat{\Psi}$ can 
differ from the unit matrix, though 
the initial value of $\hat{\psi}$ 
is equal to $1$.)  Thus we have 
the fundamental relation 
\beq
  \Psi(x,x_0,\gamma)^{-1}\hat{\Psi}(x,x_0,\kappa) 
  = \hat{\Psi}(x_0,x_0,\kappa) 
\eeq
or, equivalently, 
\beq
  \Psi(x,x_0,\gamma)^{-1}\Phi(x,x_0,\kappa) 
  \exp((x-x_0)\Lambda(\kappa)) 
  = \Phi(x_0,x_0,\kappa) 
  \label{eq:RH}
\eeq
in terms of the matrix $\Phi 
= \Phi(x,x_0,\kappa)$ introduced 
in the end of the last section.  

$\Psi$ and its inverse $\Psi^{-1}$ have 
the analytic properties mentioned in 
Sect. 3 and 4.  Namely, 
$\Psi$ have poles at $\gamma_s(x_0)$ 
and $\gamma_\infty$ with the rank-one 
structure of the residue matrices specified 
by $\bfalpha_s(x_0)$;  $\Psi^{-1}$ have 
similar properties with $\gamma_s(x_0)$ 
and $\bfalpha_s(x_0)$ replaced by 
$\gamma_s(x)$ and $\bfalpha_s(x)$.  
Having these analytic properties of 
$\Psi$, we can follow a reasoning in 
the previous paper \cite{bib:Ta03b} 
to deduce that $\Psi$ intertwines 
two ``vacua'' as 
\beq
    W_0(\bfgamma(x_0),\bfalpha(x_0)) 
  = W_0(\bfgamma(x),\bfalpha(x))
    \Psi(x,x_0,\gamma). 
  \label{eq:intertwine}
\eeq
Let us show an outline of the proof.  
A clue is the fact that $\bfalpha_s(x)$ 
is a left null vector of $\Psi$ at 
the degeneration point $\gamma_s(x)$: 
\beq
  \bfalpha_s(x)\Psi(x,x_0,\gamma_x(x)) 
  = \mathbf{0}. 
\eeq
Because of this, the poles of each element 
$\mathbf{f}$ of $W_0(\bfgamma(x),\bfalpha(x))$ 
at $\gamma_s(x)$ disappear when multiplied 
by $\Psi$.  Since $\Psi$ itself has 
simple poles at $\gamma_s(x_0)$, 
$\mathbf{f}\Psi$ in turn develops 
simple poles therein.  The rank-one structure 
of the residue matrices of $\Psi$ is 
inherited by $\mathbf{f}\Psi$.  

(\ref{eq:RH}) and (\ref{eq:intertwine}) 
imply that 
\beqnn
\lefteqn{
    W_0(\bfgamma(x),\bfalpha(x))
    \Phi(x,x_0,\kappa)} \nonumber \\
&=& W_0(\bfgamma(x_0),\bfalpha(x_0))
    \Phi(x_0,x_0,\kappa)
    \exp(-(x-x_0)\Lambda(\kappa)). 
\eeqnn
It will be more impressive to rewrite 
$\Phi(x,x_0,\kappa)$ and $\Phi(x_0,x_0,\kappa)$ 
as $\Phi(x,\kappa)$ and $\Phi(x_0,\kappa)$. 
The foregoing relation thereby reads 
\beqnn
\lefteqn{
    W_0(\bfgamma(x),\bfalpha(x))
    \Phi(x,\kappa)} \nonumber \\
&=& W_0(\bfgamma(x_0),\bfalpha(x_0))
    \Phi(x_0,\kappa)
    \exp(-(x-x_0)\Lambda(\kappa)), 
\eeqnn
which means that the ``dressed vacuum'' 
\beqnn
  W(x) = W_0(\bfgamma(x),\bfalpha(x))\Phi(x,\kappa) 
\eeqnn
obeys the exponential law 
\beq
  W(x) = W(x_0)\exp(-(x-x_0)\Lambda(\kappa)) 
\eeq
just like the usual Grassmannian perspective 
of soliton equations.  

A similar result holds for the time evolutions 
\beqnn
  \gamma_s(x) \to \gamma_s(x,t), && 
  \alpha_s(x) \to \alpha_s(x,t), \\
  \Psi(x,x_0,\gamma) \to \Psi(x,t,x_0,\gamma), &&
  \hat{\Psi}(x,x_0,\kappa) \to \hat{\Psi}(x,t,x_0,\kappa) 
\eeqnn
induced by the KP hierarchy.  Note that 
this includes the foregoing result, because 
the $t_1$-flow can be identified with 
the $x$-flow.  We omit details and show 
the final result.  

\begin{theorem} 
{\rm i)} 
$\Psi = \Psi(x,t,x_0,\gamma)$ intertwines 
two ``vacua'' as 
\beq
    W_0(\bfgamma(x_0,0),\bfalpha(x_0,0)) 
  = W_0(\bfgamma(x,t),\bfalpha(x,t))\Psi(x,t,x_0,\gamma). 
\eeq
{\rm ii)} 
The dressed vacuum 
\beqnn
  W(x,t) 
  = W_0(\bfgamma(x,t),\bfalpha(x,t))\Phi(x,t,x_0,\kappa)
\eeqnn
obeys the exponential law 
\beq
  W(x,t) 
  = W(x_0,0)\exp(-(x-x_0)\Lambda(\kappa) 
      - \sum_{\ell=1}^\infty t_\ell\Lambda(\kappa)^\ell). 
\eeq
\end{theorem}

Let us compare this result with the approach 
of Li and Mulase to commutative rings of 
ordinary differential operators 
\cite{bib:Mu90,bib:LM97,bib:Mu-review}.  
$W_0(\bfgamma,\bfalpha)$ is invariant under 
the action of the affine coordinate ring 
$\mathcal{A}$ (which is isomorphic to 
$\mathcal{A}_Q$) of the spectral curve, i.e., 
\beq
  \mathcal{A}W_0(\bfgamma,\bfalpha) 
  \subseteq W_0(\bfgamma,\bfalpha). 
\eeq
The same holds for $W(x)$ and $W(x,t)$.  
In other words, the pair $(\mathcal{A},W)$ of 
$\mathcal{A}$ and these subspaces $W$ of $V$ 
is a ``Schur pair'' in the terminology of 
Li and Mulase.   In our case, the subspace 
$W$ is parametrized more explicitly as 
$W = W_0(\bfgamma,\bfalpha)\Phi$.  
Our result shows what the KP flows 
look like in terms of the parameters 
$\gamma_s,\bfalpha_s$, etc. in this 
expression.

\subsection*{Acknowledgements} 
I would like to thank Professor Simon Ruijsenaars 
for many comments on my lecture at the workshop.  
In particular, I have to mention that I overlooked  
some important part of the paper of Carey et al. 
\cite{bib:CHMS93} and their work on free fermions 
on a Riemann surface.


\begin{thebibliography}{99}

\bibitem{bib:BC23-28}
J.L. Burchnall and T.W. Chaundy, 
Commutative ordinary differential operators, 
Proc. London Math. Soc., Ser. 2, {\bf 21} (1923), 420--440; 
Proc. Royal Soc. London, Ser. A, {\bf 118} (1928), 557--583. 

\bibitem{bib:CHMS93}
A.L. Carey, K.C. Hannabuss, L.J. Mason and M.A. Singer, 
The Landau-Lifshitz equation, elliptic curve, and 
the Ward transform, 
Commun. Math. Phys. {\bf 154} (1993), 25--47.  

\bibitem{bib:Dr77}
V.D. Drinfeld, 
On commutative subrings of some noncommutative rings, 
Funkt. Anal. Pril. {\bf 11} (1977), 11--14. 

\bibitem{bib:GD75-76}
I.M. Gelfand and L.A. Dikii, 
Asymptotic behavior of the resolvent of Sturm-Liouville equations 
and the algebra of the Korteweg-de Vries equations, 
Russian Math. Surveys {\bf 30:5} (1975), 77--113; 
The resolvent and Hamiltonian systems, 
Funct. Anal. Appl. {\bf 11} (1977), 93--105. 

\bibitem{bib:Kr77}
I.M. Krichever, 
Integration of nonlinear equations by the method of 
algebraic geometry, 
Funct. Anal. Appl. {\bf 11} (1977), 12--26; 
Methods of algebraic geometry in the theory of 
nonlinear  equations, 
Russian Math. Surveys {\bf 32:6} (1977), 185--213. 

\bibitem{bib:Kr78} 
I.M. Krichever, 
Commutative rings of ordinary linear differential operators,  
Funct. Anal. Appl. {\bf 12} (1978), no. 3, 175--185. 

\bibitem{bib:KN78}
I.M. Krichever and S.P. Novikov, 
Holomorphic vector bundles over Riemann surfaces and 
the Kadomtsev-Petviashvili equation. I,  
Funct. Anal. Appl. {\bf 12} (1978), no. 4, 276--286. 

\bibitem{bib:KN80} 
I.M. Krichever and S.P. Novikov,
Holomorphic bundles over algebraic curves, 
and nonlinear equations, 
Russian Math. Surveys {\bf 35:6} (1980), 53--80.

\bibitem{bib:Kr02}
I.M. Krichever, 
Vector bundles and Lax equations on algebraic curves, 
Commun. Math. Phys. {\bf 229} (2002), 229--269.  

\bibitem{bib:Mu90}
M. Mulase, 
Category of vector bundles on algebraic curves and 
infinite dimensional Grassmannians, 
Intern. J. Math. {\bf 1} (1990), 293--342. 

\bibitem{bib:LM97}
Y. Li and M. Mulase, 
Prym varieties and integrable systems, 
Commun. Anal. Geom. {\bf 5} (1997), 279--332.

\bibitem{bib:Mu-review}
M. Mulase, 
Algebraic theory of the KP equation, 
in R. Penner and S.T. Yau (eds.), 
``Perspectives in Mathematical Physics'' 
(International Press, 1994), 151--218.  

\bibitem{bib:Mu78} 
D. Mumford, 
An algebro-geometric construction of commutative operators 
and of solutions of the Toda lattice equations, 
Korteweg-de Vries equations and related non-linear equations, 
in ``International Symposium on Algebraic Geometry'', 
(Kinokuniya, Tokyo, 1978), 115--153 . 

\bibitem{bib:PW89}
E. Previato and G. Wilson, 
Vector bundles over curves and solutions 
of the KP equations, 
Proc. Symp. Pure Math. vol. 49, part I, 
(American Mathematical Society, 1989), 553--569 . 

\bibitem{bib:PW92}
E. Previato and G. Wilson, 
Differential operators and rank two bundles over 
elliptic curves, 
Compositio Math. 81 (1992), 107--119. 


\bibitem{bib:SS82}
M. Sato and Y. Sato, 
Soliton equations as dynamical systems on 
an infinite dimensional Grassmannian manifold, 
Lecture Notes in Num. Appl. Anal., vol. 5 
(Kinokuniya, Tokyo, 1982), 259--271 . 

\bibitem{bib:Sc05}
I. Schur, 
\"Uber verauschubare lineare Differentialausdr\"ucke, 
Sitzungsber. der Berliner Math. Gesel. {\bf 4} (1905), 2--8. 

\bibitem{bib:SW85}
G.B. Segal and G. Wilson, 
Loop groups and equations of KdV type, 
Publ. Math. IHES {\bf 61} (1985), 5--65. 

\bibitem{bib:Ta03a}
K. Takasaki, 
Tyurin parameters and elliptic analogue of 
nonlinear Schr\"odinger hierarchy, 
J. Math. Sci. Univ. Tokyo {\bf 11} (2004), 91--131. 

\bibitem{bib:Ta03b}
K. Takasaki, 
Landau-Lifshitz hierarchy and infinite dimensional 
Grassmann variety, 
Lett. Math. Phys. {\bf 67} (2004), 141--152. 

\bibitem{bib:Ta05}
K. Takasaki, 
Elliptic spectral parameter and infinite dimensional 
Grassmann variety, 
in N. Manojlovic and H. Zamtleben (eds.), 
``Infinite dimensional algebras and 
quantum integrable systems'', 
Progress in Mathematics vol. 237
(Birkh\"auser, Basel, 2005), 169--197 . 

\bibitem{bib:Ty67}
A. Tyurin, 
Classification of vector bundles over 
an algebraic curve of arbitrary genus, 
AMS Translations II, Ser. 63 
(American Mathematical Society, 1967), 245--279. . 

\bibitem{bib:Ve83}
J.-L. Verdier, 
Equations diff\'erentielles alg\'ebriques, 
S\'eminaire de l'\'Ecole Normale Sup\'erieure 1979--82 
(Birkh\"auser, 1983), 215--236. 

\end{thebibliography}
\end{document}